# The Chandra X-Ray Observatory: Progress Report and Highlights

Martin C. Weisskopf*[a]
[a]NASA/MSFC, ZP12, 320 Sparkman Drive, Huntsville, AL USA 35801

## ABSTRACT

Over the past 13 years, the Chandra X-ray Observatory's ability to provide high resolution X-ray images and spectra have established it as one of the most versatile and powerful tools for astrophysical research in the 21[st] century. Chandra explores the hot, x-ray-emitting regions of the universe, observing sources with fluxes spanning more than 10 orders of magnitude, from the X-ray brightest, Sco X-1, to the faintest sources in the Chandra Deep Field South survey. Thanks to its continuing operational life, the Chandra mission now also provides a long observing baseline which, in and of itself, is opening new research opportunities. In addition, observations in the past few years have deepened our understanding of the co-evolution of supermassive black holes and galaxies, the details of black hole accretion, the nature of dark energy and dark matter, the details of supernovae and their progenitors, the interiors of neutron stars, the evolution of massive stars, and the high-energy environment of protoplanetary nebulae and even the interaction of an exo-planet with its star. Here we update the technical status, highlight some of the scientific results, and very briefly discuss future prospects. We fully expect that the Observatory will continue to provide outstanding scientific results for many years to come.

## 1. INTRODUCTION

The Chandra X-ray Observatory, one of NASA's four Great Observatories and its flagship mission for X-ray astronomy, was launched by NASA's Space Shuttle Columbia almost exactly thirteen years ago on July 23, 1999. After the Space Transportation System launch, the observatory was boosted to high earth orbit by a separable Inertial Upper Stage followed by several burns of engines integral to the spacecraft to achieve the final orbit.

The key to Chandra's great advance in angular resolution – sub-arcsecond Full Width at Half Maximum (FWHM) – is its High-Resolution Mirror Assembly (HRMA). The 10-meter focal length HRMA has four nested pairs of cylindrical, grazing-incidence, glass-ceramic mirrors coated with iridium, the latter to enhance reflectivity at X-ray wavelengths. The eccentric Chandra orbit allows continuous observations of up to ~185 ks. The observing efficiency, currently ~75%, is limited primarily by the need to protect the instruments from particles, especially protons, during passages through the Earth's radiation belts.

The Observatory (Figure 1) consists of three main elements: (1) a telescope containing the HRMA, two X-ray transmission gratings mounted just behind the HRMA that can be inserted into the X-ray path, and a 10-meter-long optical bench; (2) a spacecraft module that provides electrical power, communications, and attitude control; and (3) the Integrated Science Instrument Module (ISIM) that holds two focal-plane cameras – the Advanced CCD Imaging Spectrometer (ACIS) and the High Resolution Camera (HRC) – and mechanisms to adjust their position. The Observatory is 13.8 m in length, 19 m in wingspan, and has a mass of 4,800 kg.

ACIS has two arrays of Charge Coupled Devices (CCDs) that provide both the position and the energy of the detected photons. The imaging array is optimized for spectrally resolved, high-resolution imaging over a 17 arcminute field-of-view; the spectroscopy array, when used in conjunction with the High Energy Transmission Grating (HETG), provides high-resolution spectroscopy with a resolving power ($E/\Delta E$) up to 1000 over the 0.4-8 keV band.

The HRC is comprised of two microchannel plate detectors, one for wide-field imaging and the other serving as readout for the Low Energy Transmission Grating (LETG). The HRC detectors have the highest spatial resolution on Chandra and, in certain operating modes, the fastest time resolution (16 μs). When operated with the HRC's spectral array, the LETG provides spectral resolution >1000 at low (0.08 – 0.2 keV) energies while covering the full Chandra energy band.

A system of gyroscopes, reaction wheels, reference lights, and a CCD-based visible light star camera enables Chandra to maneuver between targets and point stably while also providing data for accurately determining the celestial positions of observed objects. The blurring of images due to pointing uncertainty is <0.10 arcsecond, negligibly affecting the angular resolution. Absolute positions can be determined to ≤0.6 arcsecond for 90% of sources, providing an unrivaled capability for X-ray source localization.

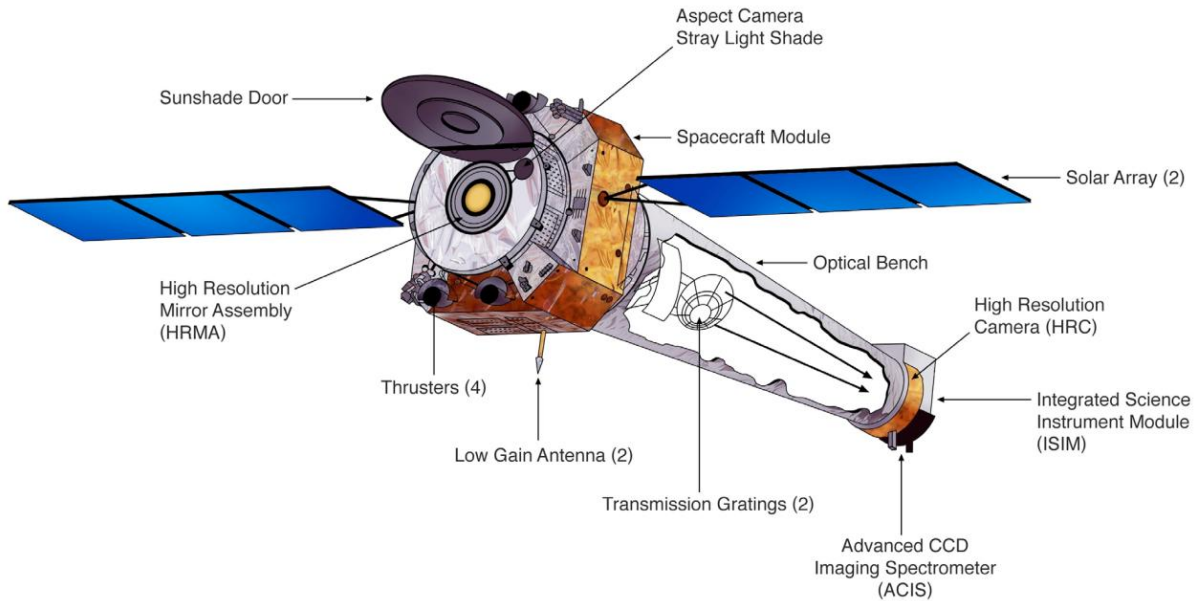

Figure 1. The Chandra X-ray Observatory.

## 2. TECHNICAL STATUS

The overall health of the observatory remains outstanding. This is most remarkable considering the observatory was designed for 3 years of operation with a goal of 5. There was, however, a successful effort not to deliberately limit the life of the mission. The observatory has experienced very few anomalies in its first 13 years of operation and there is considerable margin and robustness in the vehicle's prime systems, thus a mission life of at least 20 years is expected.

Over 13 years, of course, a few subsystems have experienced some degradation that has warranted mitigation or enhanced monitoring. For example, the vehicle's multi-layer insulation has degraded faster than expected, resulting in elevated temperatures on the Observatory's sun-facing side. The effect, due to solar particles and UV, is well characterized and is being managed with modified mission operation procedures. Overall, the insulation degradation has resulted in increased complexity planning; a reduced ability to provide uninterrupted exposure time; and, at times; a slightly lowered observing efficiency. However, the effects on scientific productivity have been minimal.

The Integrated Electron Proton Helium Instrument (IEPHIN) radiation detector has been used to ensure safing of the science instruments which then prevents damage by particles in the Earth's radiation belts or from solar outbursts. IEPHIN's temperature, however, has increased during the mission degrading its performance. We now use data from the HRC's anticoincidence shields and ACIS to accomplish this function.

The Observatory's reaction wheels are of a type that has experienced problems in a few other spacecraft after extended use however, the units are working perfectly.

To maintain pointing, the Observatory has two gyroscopes, with a total of four rotors. Standard operation uses one gyro at any time. Early in the mission, the bias current to one rotor increased, but was still below unacceptable values, perhaps indicating the possibility of future performance problems. We switched operation to the alternate unit, leaving one healthy rotor and the one with this indication of possible anomalous performance in reserve. There has been no impact to science. Incidentally, and if necessary, Chandra could be operated with one rotor from each gyro, or with only one rotor and the fine sun sensor, or even with the fine sun sensor alone.

Chandra's Aspect Camera, images guide stars in order to measure the vehicle's pointing and facilitates on-the-ground and after-the fact image reconstruction, has slowly developed a modest number of "warm" CCD pixels. The warm pixels are simply avoided when selecting locations for imaging guide stars. The rate of increase of the number of such pixels is linear in time and quite predictable. At the current growth rate, warm pixels will not impair pointing control.

All ten ACIS detector chips are functioning well, with only a few pixels per year developing into hot spots. The instrument has, however, incurred a gradual decrease in its low-energy detection efficiency due to a buildup of contamination on the optical blocking filters. The effect is greatest just above the C-K edge at 277 eV, decreasing at higher energies. A result is to limit observations that depend upon photons in the lowest energy band (<0.5 keV). Such observations require either longer exposure times or the use of the HRC or HRC plus LETG; in a very few instances Chandra simply cannot accomplish certain low-energy science observations with ACIS. The contamination is monitored every orbit using an on-board radioactive source and, occasionally, using the grating spectrometers and bright extragalactic X-ray sources. Currently the transmission of the filter is ~40% of the launch value at 0.7 keV and ~85% at 1.5 keV. Projecting forward, we predict that in five years the transmission will be 26% and 79% of launch values at these energies respectively. Even at these reduced levels, they are well within the Chandra project requirements and thus will allow science operations to proceed as at present.

Despite some early instrument anomalies, the HRC continues to operate with performance nearly unchanged since launch. Moreover, increased understanding of the imaging behavior gained with on-orbit experience has enabled improved methods of reconstructing event locations, resulting in better images. An anticipated modest decrease in the quantum efficiency (QE) of the HRC's spectral array, due to a decrease in detector gain, has been observed, and the process of raising the HRC's high voltage to recover the QE is underway.

Finally, the High and Low Energy Transmission Grating instruments are arrays of grating facets that are inserted into the beam for dispersive spectroscopic observations. These are passive elements with no discernible degradation of performance or lifetime limitations.

## 3. EXAMPLES OF RECENT SCIENTIFIC RESULTS

The Observatory is unique in its capabilities for producing sub-arcsecond X-ray images, precisely locating sources, detecting faint sources, and making high resolution, often spatially resolved spectra. The highlights presented here but a small sample drawn from the hundreds of scientific papers recently published. They illustrate the breadth and depth of research done using Chandra, as well as Chandra's importance for multiwavelength observations.

As the mission continues, new observations add to the growing archive of data, providing larger and more varied samples of objects, wider and deeper sky coverage, and deeper images and spectra. Concurrently, the aggregation of data in other wavebands increases. These increases, in turn, have improved the insights produced by theoretical models. The positive feedback generated by the increasingly rich data and knowledge base and the ingenuity of the Chandra users enhances the range and depth of scientific questions addressable. In the following we present an admittedly biased overview of some recent Chandra results. For a broader picture the reader is referred to the Chandra web site.[1]

### 3.1 Exoplanets and protplanetary disks

Chandra observations [1] of CoRoT-2, an unusual planetary system containing a hot Jupiter with an inflated radius, and a young (~200 Myr) main sequence host star (CoRoT-2a) indicate that the star is an exceptionally active X-ray emitter. The X-ray flux at the distance of the planet is roughly five orders of magnitude larger than the solar X-ray flux received by Earth and is likely eroding the planet's atmosphere. The relatively large X-ray luminosity may be a result of the star's interaction with the planet, which, in turn, could have spun the star up, or enhanced its magnetic activity thus explaining the higher X-ray activity.

---

[1] http://chandra.harvard.edu

### 3.2 Young star clusters

Chandra data were used to make a mosaic of the Great Nebula in Carina (Figure 2), yielding a catalog of >14,000 X-ray point sources, [2] more than 12,000 of which are young stars with ages between 1 and 10 Myr. Chandra's fine spatial resolution was vital in separating the point sources from the diffuse emission that pervades the region. In addition, X-ray spectra of the diffuse emission suggest that much of it is generated by charge exchange at the interfaces between Carina's hot rarefied gas and its many cold neutral pillars, ridges, and clumps.[3]

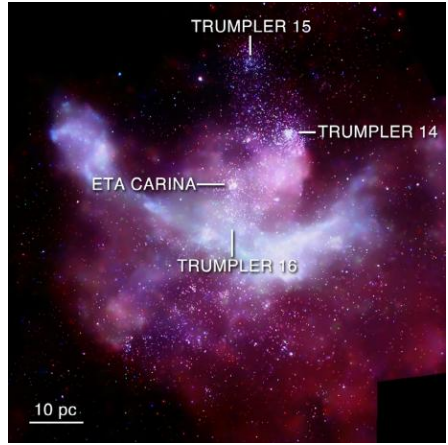

Figure 2. Diffuse X-ray emission and young stars are shown in this Chandra image [2] of the Carina Nebula. The survey of the Nebula covers a field of 1.4 deg$^2$ and is constructed from a mosaic of 22 individual observations. The positions of Eta Carinae and three star clusters are labeled.

### 3.3 Stellar winds

Stellar winds carry away a significant portion of a massive star's mass during its evolution, and are a source for the deposition of energy, momentum and matter into the interstellar medium. X-ray emission-line profile analysis provides an independent way to measure the mass-loss rates of such stars. Unlike ultraviolet (UV) absorption line diagnostics, X-ray profile analysis is not very sensitive to the ionization balance and it relies on continuum rather than line opacity so the analysis is not subject to the uncertainty associated with saturated absorption.

Recent Chandra HETG observations of two O supergiants, ζ Pup and HD 93129A [4, 5] (Figure 3), have led to a reevaluation of the mass-loss rates for these stars. For both, the X-ray emission-line profile modeling indicates mass-loss rates 3-4 times lower than those determined from smooth-wind models, but consistent with estimates which include small-scale clumping, hence refining our understanding of these processes and systems. The low values for the mass-loss rates are also confirmed by independent measurement of X-ray continuum absorption. These types of studies are now being extended to other massive stars.

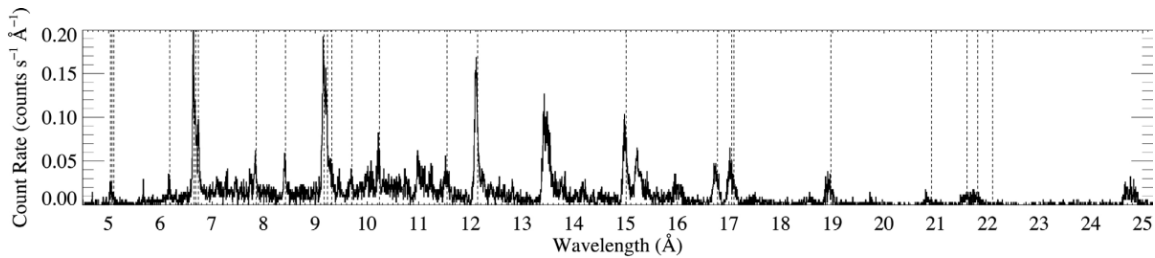

Figure 3. A Chandra HETG spectrum of ζ Pup [5]. The most prominent lines here are the He-like complexes from Si XIII (6.65, 6.69, & 6.74 Å) and Mg XI (9.17, 9. 23, & 9.31 Å), Ly$_\alpha$ lines from Ne X (12.13 Å), OVIII (18.97 Å), and Fe XVII lines (15.01, 16.78, 17.05, & 17.10 Å). Vertical dashed lines show the laboratory rest wavelengths of lines fit with a wind profile model. The asymmetry in the line profiles is a consequence of the moderate mass-loss rate of 3.5±0.3 10$^{-5}$ M$_\odot$/ yr.

## 3.4 Neutron stars and matter at extreme densities

Neutron star (NS) cores contain super dense matter whose detailed physical properties are still uncertain. NSs are heated to billions of degrees during the supernova process that creates them, and then they cool via a combination of neutrino and photon emission. Observing the cooling rates of young NSs then offers a method for studying the behavior of matter at extreme densities. Ten years of Chandra data (Figure 4) show a 21% decrease in the flux of the NS in the 330-year old Cas A supernova remnant. The absence of pulsations from the NS indicates that the emission is probably from the entire surface. It has been shown [6] that the spectrum can be fit to a non-magnetized (B $<10^{11}$ G) carbon atmosphere model with a NS mass M = 1.65 $M_\odot$, and radius R = 10.3 km. The spectral data imply a relative decline in surface temperature of 4% (5.4σ) over a ten-year period [7], consistent with what is observed. The temperature decline is too rapid to be caused by normal thermal cooling processes, but can be explained if neutrons in the core have recently undergone a transition to a superfluid state. If confirmed, this result provides the first direct evidence that superfluidity and superconductivity occur at supranuclear densities within NSs [8, 9] and provides an important probe of the nuclear strong force.

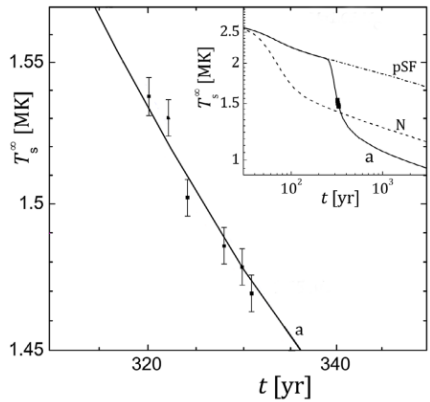

Figure 4. The temperature decline of the NS in the Cas A SNR observed with Chandra [8], along with model predictions. The solid line shows cooling predicted by the best-fit model: a 1.65 $M_\odot$ NS of age 330 years with strong proton superfluidity and moderate neutron superfluidity. The inset shows the cooling predicted by this model ("a") over a longer time span. Also shown in the inset are predictions of models with only proton superfluidity ("pSF") and a model with neither proton nor neutron superfluidity ("N").

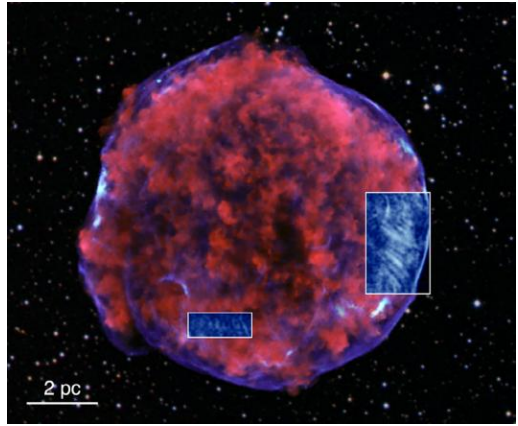

Figure 5. A Chandra image [13] of the Tycho SNR, showing ejecta in red (low energy X-rays) and the forward shock in blue (high energy X-rays). The two boxes show only high energy X-rays, and reveal a pattern of X-ray stripes never previously seen in a SNR.

### 3.5 Supernova remnants as sites of particle acceleration

Energetic arguments support the hypothesis that acceleration of particles in supernova shock waves are the source of cosmic rays (mostly protons and helium nuclei) up to the knee in the cosmic ray spectrum at $10^{15}$ eV [10]. The recent detection [11] of 100 GeV gamma rays from the Tycho SNR by the Fermi space telescope supports this model, but an explanation of the γ-ray data in terms of energetic electrons rather than protons is also possible [12]. These possibilities are addressed by deep new Chandra observations of the SNR which discovered strikingly ordered patterns of stripes (Figure 5) with non-thermal spectra. The spacing between the stripes is found to correspond to the gyroradii of $10^{14}$ to $10^{15}$ eV protons [13] for magnetic field strengths of a few to a few tens of microgauss. This result is consistent with a recently developed model [14] for shock acceleration in which a cosmic-ray current driven instability can amplify the magnetic field and produce narrow peaks in the magnetic turbulence with a separation ~ a gyroradius. Therefore, the X-ray stripes observed in Tycho may provide direct evidence for acceleration of cosmic rays up to $10^{15}$ eV.

### 3.6 Feedback in groups and clusters of galaxies

There are enormous reservoirs of hot, x-ray-emitting gas in clusters of galaxies that contain a record of past activity (Figure 6): giant cavities from explosions hundreds of millions of years ago, shock fronts from past mergers, and trace elements that reflect transport of supernova-enriched gas out of the galaxy into the intracluster medium (ICM). Moreover, through detailed observations with Chandra, one is able to show that the fraction of shock energy that goes into heating the gas (5-10%) is sufficient to balance radiative cooling locally at the shock fronts and the outburst interval is short enough for such shocks to offset cooling over much longer timescales. This demonstrates that the phenomenon we refer to as AGN feedback can operate to heat the gas within galaxies, as well as the more extended ICM.

Deep (574 ks) Chandra observations of the giant elliptical galaxy M87 in the Virgo cluster ([15]; Figure 6-left ) and the Abell 2052 cluster ([16]; Figure 6-right) verify this picture of compressed bubble rims, filaments and loops providing evidence for heating by repeated shocks occurring at intervals ~10 Myr.

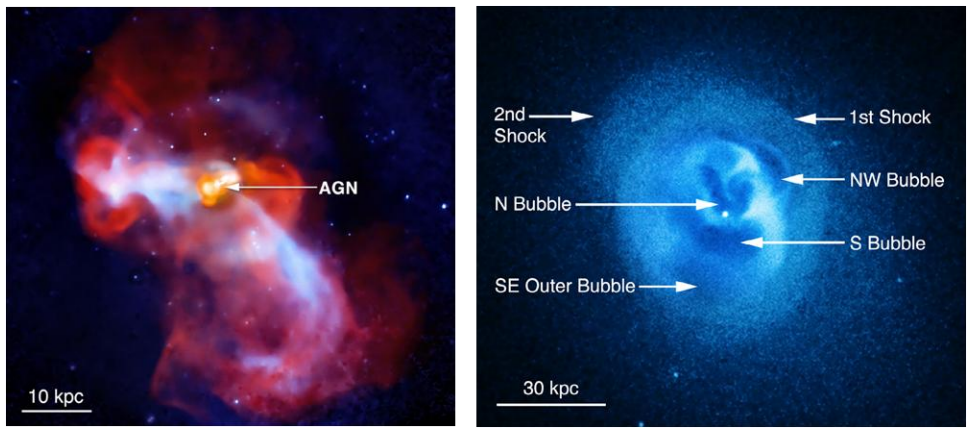

Figure 6. Left. A composite image of M87 combining data from Chandra (blue) and the VLA (red). The overlapping X-ray and radio emission on the eastern (left) side of the nucleus shows radio-emitting plasma entraining and lifting cooler, metal-enriched gas [15]. Right. Chandra image [16] of the central region of the galaxy cluster Abell 2052, showing a complicated set of structures including multiple bubbles and two shock fronts resulting from AGN feedback.

### 3.7 Constraints on cosmological models and dark energy

The standard Lambda Cold Dark Matter (ΛCDM) cosmological model is based on a bottom-up sequence of structure formation, largely dominated by the gravity of the cold dark matter. This leads to picture where there is a hierarchical series of mergers of galaxies, then groups and small clusters of galaxies, and culminates in the formation of clusters of galaxies. Large fluctuations, and correspondingly large cluster masses, should therefore be extremely rare in the early universe. The space density of massive, high-redshift clusters is thus a particularly useful measure for testing this picture and Chandra observations of the mass function of clusters have been used to confirm the existence of dark energy and to constrain cosmological parameters such as the energy density $\Omega_\Lambda$ and matter density $\Omega_m$, which characterize the dark energy and matter densities [17], and to test modified gravity models designed to explain the acceleration of cosmic expansion [18].

Chandra has also played a key role in the implementation of a powerful observational technique that combines X-ray observations with data from a new generation of dedicated millimeter-wave telescopes, including the Atacama Cosmology Telescope (ACT), the South Pole Telescope (SPT), and the Planck space observatory, to find and study massive clusters of galaxies. The technique exploits the Sunyaev-Zeldovich (SZ) effect [19], a distortion in the cosmic microwave background (CMB) spectrum caused by Compton scattering of CMB photons with the hot intracluster medium. The SZ effect is independent of cluster redshift and is tightly correlated with the cluster mass as determined by X-ray observations. The accuracy with which the observed mass proxies can be linked to the true cluster mass places a fundamental limit on the precision of cosmological constraints. An example of the importance of Chandra for calibrating SZ surveys is illustrated by a recent study of a sample of 15 clusters observed by the SPT in the redshift range $0.29 < z < 1.08$ [20].

Combining X-ray measures with Chandra of the growth rate of clusters with redshift [21], X-ray measures with Chandra of the gas mass fraction in clusters [22], current cosmic microwave background (CMB) measurements with NASA's Wilkinson Microwave Anisotropy Probe (WMAP) [23], supernova type 1a [24] and baryon acoustic oscillation (BAO) data [25] constrains cosmological parameters: the matter density $\Omega_m = 0.27\pm0.02$, $\sigma 8 = 0.79\pm0.03$ and the equation of state $w = -0.96\pm0.06$ for a particular set of models --- spatially flat with a constant dark energy equation of state parameter w (Figure 7). A value of $w = -1$ corresponds to dark energy being described by a cosmological constant term in General Relativity.

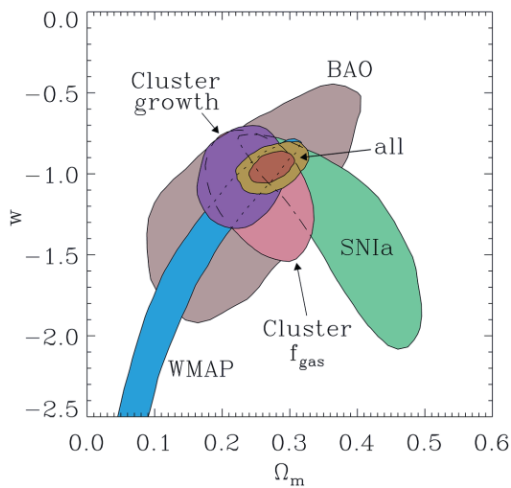

Figure 7. Constraints on the cosmological parameters w and $\Omega_m$ obtained using different data sets. Joint 68.3% confidence regions are shown for examining the growth of clusters [21], for a technique that also made use of Chandra assumes the X-ray emitting gas fraction is constant independent of redshift [22], 5-yr WMAP data [23], supernova Ia data [24], and Broad Acoustic Oscillation (BAO) data [25]. 68.3% and 95.4% confidence regions, when combining the five data sets, are shown in red and gold respectively.

## 4. LOOKING TO THE FUTURE

The exact targets for observations in the future will always be determined by peer review as they are now. Past experience tells us that the capabilities of Chandra and the seemingly inexhaustible complexity of the cosmos will provide major surprises in what those observations reveal. We have highlighted several examples from the past few years and we confidently anticipate that new Chandra observing programs will build on the highly successful synergies with other observatories, an example being a potential collaboration on a new Hubble Ultra Deep Field. Joint Chandra projects with the Atacama Large Millimeter Array (ALMA) will facilitate new studies of the influence of X-radiation on protoplanetary disks, the evolution of obscured supermassive black holes, and many other phenomena with complementary sub-arcsecond resolution in both sub-mm and X-ray wavelengths

Over the next several years, observations of Cas A, spaced by approximately one year, will continue to monitor the surface temperature evolution of the NS and test the model described in the section 3.4 which predicts a continued decline in temperature.

It is safe to say that future observations with Chandra will also further probe cosmic ray acceleration processes. For example, SNR G1.9+0.3 is the only Galactic SNR that is known to be brightening at both X-ray and radio wavelengths. This brightening implies that we are observing particle acceleration in real time. Detecting spectral changes with Chandra as the SNR brightens in the future should give an indication of electron synchrotron lifetimes and energies, and thus provide more input for the theory of shock acceleration.

Perhaps one of the most compelling examples for future Chandra observation follows from the deepest Chandra image (Figure 8). Despite its depth, the image remains photon- limited rather than source- confusion- or background-limited implying that a major increase in observing time would yield exciting new results.

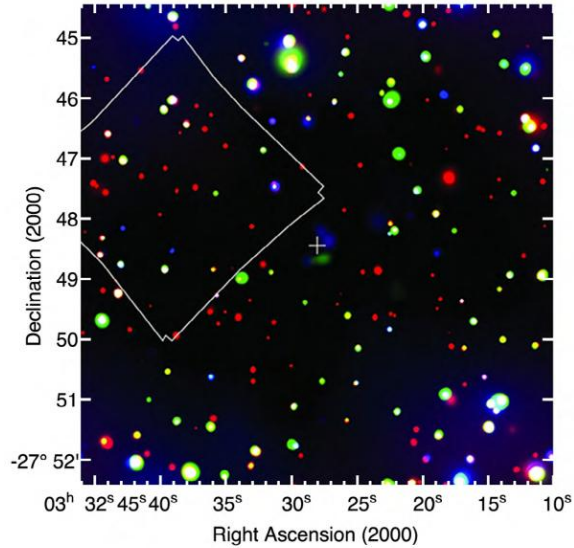

Figure 8. A Chandra image of the inner 8' x 8' of Chandra Deep Field South. This is the deepest X-ray image currently available and is a composite of smoothed images in the 0.5-2.0 keV (red), 2-4 keV (green) and 4-8 keV (blue bands. The polygon shows the field of view of the Hubble Ultra Deep Field [26].

## ACKNOWLEDGEMTS

As Chandra Project scientist, I have relied heavily on previous papers, reports, presentations, proposals to NASA's Senior Review, and other documents that I and my colleagues have prepared on behalf of the Chandra Mission. I especially want to acknowledge the direct and indirect contributions of the entire Chandra team with special thanks to W. Tucker, P. Edmonds, H. Tananbaum, R. Brissendon and my deputy S. O'Dell.